\documentclass[aps, pra, reprint, superscriptaddress,nofootinbib]{revtex4-1}
\usepackage[charter]{mathdesign}
\usepackage{amsmath,graphicx,bm}
\usepackage[plainpages=false,colorlinks=true,linkcolor=blue, citecolor=blue, urlcolor=blue]{hyperref}
\usepackage{color}
\usepackage{physics}
\usepackage{comment}
\usepackage{graphicx} 
\usepackage{dsfont}

\def\ket#1{| #1 \rangle}

\def\bracket#1#2{\langle #1 | #2 \rangle}
\def\kb#1#2{| #1 \rangle\!\langle #2 |}

\def\cN{\mathcal{N}}

\def\fig#1{Fig.~\ref{fig:#1}}

\newcommand{\nearest}[2]{\left\langle #1,\, #2\right\rangle}

\begin{document}

\setlength{\abovedisplayskip}{2pt}
\setlength{\belowdisplayskip}{2pt}
\setlength{\abovedisplayshortskip}{2pt}
\setlength{\belowdisplayshortskip}{2pt}

\title{Resource estimate for quantum many-body ground-state preparation on a quantum computer}

\author{Jessica Lemieux}
\affiliation{D\'epartement de Physique \& Institut Quantique, Universit\'e de Sherbrooke, Qu\'ebec, Canada J1K 2R1}
\author{Guillaume Duclos-Cianci}
\affiliation{D\'epartement de Physique \& Institut Quantique, Universit\'e de Sherbrooke, Qu\'ebec, Canada J1K 2R1}
\author{David S\'en\'echal}
\affiliation{D\'epartement de Physique \& Institut Quantique, Universit\'e de Sherbrooke, Qu\'ebec, Canada J1K 2R1}
\author{David Poulin}
\affiliation{D\'epartement de Physique \& Institut Quantique, Universit\'e de Sherbrooke, Qu\'ebec, Canada J1K 2R1}
\affiliation{Canadian Institute for Advanced Research, Toronto, Ontario, Canada M5G 1Z8}
\affiliation{Quantum Architecture and Computation Group, Microsoft Research, Redmond, WA 98052, USA}
\date{\today}

\begin{abstract}
	We estimate the resources required to prepare the ground state of a quantum many-body system on a quantum computer of intermediate size. This estimate is made possible using a combination of quantum many-body methods and analytic upper bounds. Our routine can also be used to optimize certain design parameters for specific problem instances. Lastly, we propose and benchmark an improved quantum state preparation procedure. We find that it reduces the circuit T-depth by a factor as large as $10^6$ for intermediate-size lattices. 
\end{abstract}

\maketitle

\section{Introduction}

Quantum many-body systems are notoriously difficult to simulate on classical computers. This observation led Feynman to suggest that they could be efficiently simulated using quantum computers instead~\cite{Feynman1982}. The {\em dynamics} of quantum systems are dictated by Schr\"odinger's equation, which results in a unitary time evolution. Beginning with Lloyd~\cite{Lloyd1073} and Aharonov and Ta-Shma~\cite{AT03b}, the design of increasingly efficient quantum circuits solving Schr\"odinger's equation has been an intensive area of research in the past decade (see, e.g., Refs.~\cite{CS19a,RSKS18a,CROD18a,PHWW14a,WBCH13a,JLP11a,RWS12a,BC12a,BCCK15a,BBKW16a,C18a,LC17a}).

Beyond the dynamics, the simulation of a quantum system also requires setting initial conditions. Of particular interest are properties at low temperature, where, to a good approximation, the system is initially in its ground state. Thus, to solve the {\em static} problem, a quantum circuit must be constructed that maps a fiducial initial state to the ground state of the system of interest. This problem is generally QMA-complete~\cite{KSV02a,KKR06a,KR03a,OT05c}, so the existence of a general-purpose efficient procedure is believed to be impossible. Nonetheless, heuristic methods have been proposed~\cite{AT03b,BKS09a,TOVP11a,YA12a,KGBW19a} that could be efficient for specific physical systems, and exponential-time algorithms~\cite{PW09a,WBA10a} could be sufficiently fast for intermediate-size problems. 

The situation is somewhat reversed when it comes to classical computers. There, a host of methods have been devised to approximately solve the static problem. These include density-functional theory, quantum Monte Carlo, and tensor-network methods, to name a few (see, e.g., Refs.~\cite{Jones2015,Acioli1997,Orus2014} for reviews). However, the growth of entanglement in time typically makes it impossible to classically simulate the dynamics of quantum many-body systems.

The goal of this article is to compare and optimize various approaches to provide an estimate of the resources (number of gates) required to solve a quantum simulation problem on an intermediate-size quantum computer, say of a few hundred qubits. For the sake of concreteness, we study the Hubbard model, the paradigmatic model of strongly correlated electrons. We will also focus on a specific, generally nonefficient algorithm to prepare the ground state of this model; it is a discrete version~\cite{BKS09a} of the quantum adiabatic state preparation~\cite{AT03b,Farhi2001,Farhi2000} with basic components that have also been presented in Ref.~\cite{Berry2018}. For simplicity, we consider a linear interpolation with an optimized schedule. The latter provides a rough upper bound that can surely be improved, e.g. by adding a symmetry-breaking field to lift gapless modes~\cite{WHWC15a}. The runtime of this algorithm depends on properties of the system usually unknown analytically. We employ tensor-network methods to (approximately) solve the static problem on a classical computer. Aside being useful to benchmark the quantum algorithm, this classical side computation could also be used to optimize certain parameters of the adiabatic state preparation algorithm, such as deciding on an interpolation schedule. This idea of combining classical and quantum algorithm is often seen in Quantum Approximate Optimization Algorithm (QAOA), for example~\cite{farhi2014quantum}.

\section{The Hubbard model} 
The Hubbard model is defined by the Hamiltonian 
\begin{equation}
	H(T_\text{hub},U)= T_\text{hub}\sum_{\sigma \, \nearest ij}^{} (c_{i\sigma}^\dagger c_{j\sigma} +c_{j\sigma}^\dagger c_{i\sigma})+U\sum_{i}^{}n_{i\uparrow} n_{i\downarrow}
\end{equation}
where the operator $c_{i\sigma}^\dagger$ creates an electron of spin $\sigma =\{\uparrow,\downarrow\}$ at site $i$, its adjoint $c_{i\sigma}$ is the corresponding annihilation operator, and $n_{i\sigma} = c_{i\sigma}^\dagger c_{i\sigma}$ is the number operator. The notation $\langle i,j\rangle$ indicates nearest-neighbor sites on a lattice. The constant $T_\text{hub}$ in the kinetic term defines the energy unit and is henceforth set to $T_\text{hub}=1$. $U$ is the strength of the Coulomb interaction, limited to electrons on the same site.

The kinetic term tends to delocalize electrons, and indeed the limit $U/T_\text{hub}=0$ is easily solved as a free Fermi gas. 
At the other extreme ($T_\text{hub}/U=0$), the electrons are perfectly localized on each site at half-filling.
In between these two extremes, the kinetic and Coulomb energies are in competition and the solution is highly nontrivial.
In fact, in the thermodynamic limit, the ground-state phase diagram of the two-dimensional Hubbard model is expected to feature various broken-symmetry states (antiferromagnetic, superconducting, etc) depending on the electron density, with a vanishing energy gap.

In this work, we set the value of $U$ to be twice the number of neighbors -- e.g., $ U = 4 $ for a one-dimensional (1D) chain, $ U = 6 $ for a ladder and $ U = 8 $ for a two-dimensional (2D) rectangular lattice -- and consider a 10\% electron doping above half-filling, with an approximately equal number of up and down spins. For small system sizes, we found that this doping has the smallest gap at fixed number of spins. Therefore, with these parameters, the systems should have a small gap along the adiabatic path chosen, which defines hard instances for the state preparation algorithm.

\section{Discrete adiabatic state preparation} 

The adiabatic algorithm~\cite{Farhi2001,Farhi2000} leverages a quantum computer's ability to simulate Schr\"odinger's equation in order to prepare the ground state of a target Hamiltonian $H(\tau)$ using the adiabatic theorem. This is made possible by beginning with a Hamiltonian $H(0)$ with a known and easy-to-prepare ground state, and slowly morphing it into the Hamiltonian of interest $H(\tau)$. The algorithm is efficient if $H(t)$ has a gap, $\Delta(t)=E_1(t)-E_0(t)$, polynomial in the system size at all times $0\leq t\leq \tau$. 

In this work, we will use a discrete version of the adiabatic algorithm~\cite{BKS09a}, reminiscent of the quantum Zeno effect. Recall that the latter employs projective measurements to freeze the unitary evolution of a quantum system. Here, we instead use a sequence of time-dependent measurements to drag the state of a quantum system that is otherwise static. If the change in the measurement basis is small at every step, the outcome is almost deterministic. 

Concretely, we choose a discrete sequence of Hamiltonians $H_j$, $j=0,1,\ldots , L$ with corresponding ground states $\ket{\psi_j^0}$. Given the fidelity $F_j = |\bracket{\psi_{j-1}^0}{\psi_{j}^0}|^2$ between two consecutive ground states, we can express 
\begin{align*}
	\ket{\psi_{j-1}^0} = \sqrt{F_j} \ket{\psi_j^0} + \sqrt{1-F_j}\ket{\overline{\psi_j^0}},
\end{align*}
where $\ket{\overline {\psi_j^0}}$ denotes a state orthogonal to the ground state of $H_j$. We define a binary projective measurement by the projectors $Q_j = \kb{\psi_j^0}{\psi_j^0}$ and $\bar Q_j = I-Q_j$. If the system is in state $\ket{\psi_{j-1}^0}$, then a measurement of $\{ Q_j, \bar Q_j \}$ will produce the outcome $Q_j$ with probability $F_j$, causing the state to collapse into $\ket{\psi_j^0}$. Successively measuring for $j=1$ to $L$ will produce the desired state $\ket{\psi_L^0}$ with a global success probability $p = \prod_{j=1}^L F_j$. The main interest of using projective measurements instead of adiabatic evolution is that errors do not accumulate. The error is limited to the one of the last projective measurement.

The binary measurement $\{Q_j, \bar Q_j\}$ is a coarse-grained energy measurement and can in principle be performed from quantum phase estimation~\cite{Kit95a,CEMM98a,AL99a} using the operator $U_j = e^{-iH_j}$. Recall that this algorithm is an effective way of estimating the eigenvalues $e^{i\phi_k}$ of a unitary operator $U$. Thus, to ensure that the quantum phase estimation algorithm differentiates all eigenstates, we normalize $H_j$ such that its eigenvalues are between $0$ and $2\pi$. A measurement accuracy $\epsilon$ requires a simulation time of $O{( 1/\epsilon)}$. In the present setting, we are interested in distinguishing the ground state of $H_j$ from the rest of the spectrum. Based on the above properties of quantum phase estimation, the binary measurement $\{Q_j, \bar Q_j\}$ requires in general simulating the time evolution under $H_j$ for a time $t_j = 1/\Delta_j$, where $\Delta_j$ is the spectral gap of $H_j$. This can be though of as a manifestation of the Heisenberg time-energy uncertainly relation. 

Note that, because nondestructive energy measurements are central to many quantum algorithms, other methods have been devised recently to realize them~\cite{DSHT17a}. Since the relation between the complexities of these methods is relatively well understood, our results can be easily translated to any of these alternative methods by replacing $t_j$ by the appropriate complexity metric. Whether based on a simulation of time evolution or otherwise, in all cases, a circuit implementation of the binary measurement $\{Q_j, \bar Q_j\}$ requires knowing the ground-state energy $E_j^0$ and the spectral gap $\Delta_j$. We will return to this requirement when describing the simulation method. 

In the following, we will estimate the complexity of state preparation in terms of the total simulation time. For a given Hamiltonian sequence $\{H_j\}_{j=0}^L$, the duration of a successful state preparation is simply 
\begin{equation}
t_{\rm total} = \sum_{j=1}^L t_j = \sum_{j=1}^L \frac 1{\Delta_j}.
\end{equation}
The algorithm succeeds with probability $p$. To achieve any desired overall success probability $1-\epsilon$, it needs to be repeated $\log(\epsilon)/\log(1-p)$ times, resulting in a total time to solution (TTS)~\cite{Ronnow2014} of
\begin{equation}
{\rm TTS}(\epsilon, \{H_j\}) = \frac{\log(\epsilon)}{\log(1-\prod_{j=1}^L F_j)}  \sum_{j=1}^L \frac 1{\Delta_j}.
\label{eq:TTS1}
\end{equation}

The TTS needs to be minimized over some Hamiltonian sequences, achieved by the following procedure: We initialize the sequence with $\{H_0, H_1\}$ where $H_0$ is the simple Hamiltonian ($U=0$ in our case) and $H_1$ is the Hamiltonian of interest. At iteration $L$, we have the sequence $\{H_j\}_{j=0}^L$ and the corresponding fidelities $\{F_j\}_{j=1}^L$, and for $k= {\rm argmin}_j F_j$, we perform the assignation 
\begin{align}
H_{j} &\leftarrow H_j \quad {\rm for} \ j=0\ldots k-1 \\  
H_{k} &\leftarrow (H_{k-1}+H_k)/2 \\  
H_{j+1} &\leftarrow H_j \quad {\rm for} \ j=k\ldots L+1. 
\end{align}
In other words, we add a Hamiltonian halfway between the two Hamiltonians with the lowest fidelity in the sequence. We end the procedure when ${\rm TTS}(\epsilon, \{H_j\})$ has stopped decreasing for a few iterations, i.e., when we are convinced that the algorithm found the minimum. The latter defines the optimal trade-off between space and time cost: Few repetitions of a long high-fidelity iteration or several shorter iterations of lower fidelity.  

\section{Rewind procedure}

The above procedure has a finite probability of failure, and in case of failure the whole state preparation must be restarted from scratch. Here, we propose a modification to the algorithm which avoids such hard reboots. The technique was introduced in Ref.~\cite{MW05a} and used in Refs.~\cite{TOVP11a,WHWC15a}. 

At step $j$, the register is in state $\ket{\psi_{j-1}^0}$ and the measurement $\{Q_j, \bar Q_j\}$ is performed. The outcome $Q_j$ (resp. $\bar Q_j$) occurs with probability $F_j$ (resp. $1-F_j$) and yields the state $\ket{\psi_j^0}$ (resp. $\ket{\overline {\psi_j^0}}$). Instead of rejecting this outcome and rebooting to $j=0$, we alternate measurements of $\{Q_{j-1}, \bar Q_{j-1}\}$ and $\{Q_j, \bar Q_j\}$, and halt whenever $Q_j$ occurs. 

\begin{figure}[ht]
\centering
\includegraphics[width=0.5\linewidth]{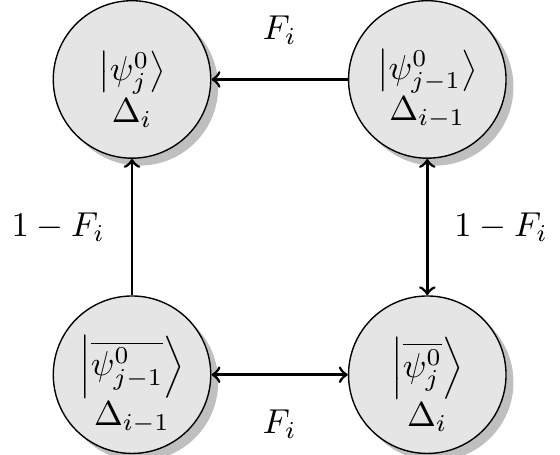}
\caption{The cost of the rewind procedure for one step will be the sum of all paths that start at the top right state and end at the top left one. The probability to follow an arrow is given by the fidelity $F$ and its cost is given by the gap $\Delta$ where the arrow ends.}
\label{fig:TTSr}
\end{figure}

It can easily be shown~\cite{MW05a} that the probability of a $Q_{j-1} \leftrightarrow Q_j$ or a $\bar Q_{j-1} \leftrightarrow \bar Q_j$ transition is $F_j$, while the other transitions $Q_{j-1} \leftrightarrow \bar Q_j$ and $\bar Q_{j-1} \leftrightarrow Q_j$ have complementary probabilities $1-F_j$. Thus, each step in the Hamiltonian sequence can be seen as a random walk on the graph depicted on \fig{TTSr}, with initial condition $\ket{\psi_{j-1}^0}$ and absorbing state $\ket{\psi_j^0}$. The cost of each transition is given by the inverse of the spectral gap $\Delta$ of the target state. These considerations lead to a simple geometric series for the average time $A_j$ to realize step $j$:
\begin{equation}\label{eq:TTS2}
	\small A_j = \frac{F_j}{\Delta_J} + 2 \sum_{k_1=1}^\infty \sum_{k_2=0}^{k_1}(1-F_j)^{2k_1} F_j^{2k_2+1}\Big(\frac{k_1+k_2+1}{\Delta_j} + \frac{k_1+k_2}{\Delta_{j-1}}\Big).
\end{equation}
The average of the TTS is TTS$(\{H_j\}) = \sum_{j-1}^L A_j$ and can be minimized over Hamiltonian sequences $\{H_j\}$ following the procedure described in the previous section. 

\section{Qubitization}

The preparation cost is a function of the gap and path of evolution, but the complexity of the whole preparation also depends on the implementation of the time-evolution operator. In Ref.~\cite{DSHT17a}, a unitary walk operator $W=\exp{i \acos(\bar Ht)}$ is introduced with an efficient implementation. The spectrum of the Hamiltonian is reversed and renormalized,  $\bar H_j =  \mathbb{I}-{H_j}/{\cN_j} = \sum_l |\beta_l^j|^2 P_l$, where $\cN_j$ is the normalization factor and the $P_l$ are Pauli operators. Note that, since we considered the spectrum of ${H_j}$ to be between $0$ and $2 \pi$, $\cN_j =  2\pi$ in the current case. The walk operator maps the Hamiltonian to a higher-dimensional space, $W=SVe^{i\pi}$ where we defined
\begin{align}
	B\ket{0} &= \sum_j \beta_j \ket{j}, \\
	S &=B(1-2\ketbra{0})B^\dagger, \text{ and } \\
	V &= \sum_j \ketbra{j} P_j
\end{align}
as in Refs.~\cite{BCCK15a,Low2019,DSHT17a}.

The algorithm described in the previous sections uses a discretization of the time-dependent Hamiltonian 
\begin{equation}\label{eq:Ht}	
	H(t) = \bigg(1- \frac{t}{\tau} \bigg)H_0 + \frac{t}{\tau} H_f,
\end{equation}
where $\tau$ is the adiabatic parameter, $H_0$ is the initial Hamiltonian with an easy-to-prepare ground state, and $H_f$ is the final Hamiltonian with the target ground state. Recall that energy measurement, such as quantum phase estimation, uses the unitary $e^{iH_j}$ to decompose the current state in the $k$ eigenstates of $H_j$,
\begin{equation}
QPE \ket{0}\ket{\psi_{j-1}^0} = \sum_k \braket{\psi_j^k}{\psi_j^0}\ket{E_j^k}\ket{\psi_j^k},
\end{equation}
the algorithm mimics the evolution by partial energy measurements in order to drag the state towards the ground state of the next instantaneous Hamiltonian.  

When adding qubitization, the initial state is the corresponding eigenstate of $W_0$, $\ket{\varphi_0^0}$. The eigenstates of $W_j$ are given by
\begin{align*}
\ket{\varphi^k_j} &= \frac{1}{\sqrt{2}} \Bigg ( \Big[1 \mp \frac{i \bar E^k_j}{\sqrt{1-(\bar E^k_j)^2}}\Big] \sum_l \beta_l^j \ket{l}\ket{\psi^k_j} \\
	& \qquad \pm \frac{i}{\sqrt{1-(\bar E^k_j)^2}} \sum_l \beta_l^j (\mathds{1} \otimes P_l) \ket{l}\ket{\psi^k_j} \Bigg ).
\end{align*}
The qubitized algorithm uses the same discretization $\{H_j\}$, but with the corresponding unitary $W_j=\exp{i \acos(\bar H_j)}$ for the partial energy measurement. The states will then go from $\ket{\varphi_{j-1}^0}$ to $\ket{\varphi_{j}^0}$. Note that it does not correspond to any meaningful physical evolution. Thus, performing phase estimation of $W_j$ instead of $e^{iH_j}$ will lead to the preparation of a state equivalent to the ground state of the Hubbard model, up to an isometry. In addition, the implementation of $W_j$ is exact and, in general, costs less than the implementation of $e^{iH_j}$, because $W_j$ can be implemented with approximately the same amount of resources needed in a single Trotter step~\cite{DSHT17a}.

If needed, at the end of the algorithm, the final state is mapped back to the lower-dimensional space with a known isometry. However, some properties such as any static expectation values, can be derived directly from $\ket{\varphi_f^0}$, but since $W$ is not a physical Hamiltonian, we do not expect its dynamic to behave as $H_f$. 

In the usual adiabatic setting, we assume that this time-dependent Hamiltonian of Eq.(\ref{eq:Ht}) has a nonvanishing gap for $0 \leq t  \leq \tau$ and the scaling is bounded by $O\big(1/{\Delta^2}\big)$~\cite{959902}. Intuitively, the scaling can be understood as follows:  Each implementation of the QPE needs to differentiate the ground state of the first-excited state, so the required precision of the energy scales with $1/{\Delta}$. The density of steps, i.e., the discretization, depends on how fast the time-dependent Hamiltonian spectrum changes, which can be partially characterized by the gap.

With the qubitization protocol, we have a degeneracy of the states but it will not affect the protocol since both $k$th eigenstates of $W_j$ correspond to eigenstate $k$ of $H_j$. The relation to the gap of interest, the one that differentiates $k=0$ from $k=1$ is given by
\begin{equation}\label{eq:Wgap}
\Delta_{W_j} = \acos\bigg(1-\frac{\Delta}{\cN_j}\bigg).
\end{equation}
 In some cases, the gap may decrease, which is not good for our purpose. However, for small gaps, in many cases, we will have a significant advantage since
\begin{equation}
 \Delta_{W_j} = \acos\bigg(1-\frac{\Delta}{2\pi}\bigg) > \Delta   \text{ for } \Delta \lessapprox 0.32
\end{equation}
To summarize, the discretization should still be bound by $1/{\Delta}$, but the QPE will now scale with $1/{\Delta_{W_j}}$.

For more details on qubitization applied in this context, see Ref.~\cite{Berry2018}. 

Qubitization affects the success probability and the gap in an instance-dependent manner. Thus, we cannot formally evaluate its impact on the cost. To assess the efficiency of the algorithm, we performed numerical simulations.

\section{Simulation method}

The resources required for a given state preparation depend on the gap of every Hamiltonian in the sequence and the fidelity between the ground state of two consecutive Hamiltonians. For small lattices, these can be computed by Lanczos exact diagonalization~\cite{lanczos1950iteration}. However, this method becomes intractable for system sizes of up to a hundred qubits. To deal with this issue, we use the density-matrix renormalization group (DMRG, see Ref.~\cite{S11b} for a review), a tensor network method that represents quantum states by one-dimensional networks known as matrix product states (MPSs). DMRG is thus the method of choice to study one-dimensional systems numerically, or two-dimensional systems of small width~\cite{Stoudenmire2012}. The method has been implemented using the ITensor C++ library~\cite{ITensor}. While the results are approximate, we validate the small-size instances with exact diagonalization. 

This simulation method is not only of interest for classically benchmarking the quantum state preparation algorithm but, more broadly, can be used to optimize the Hamiltonian sequence. In addition, the circuit implementation of the projective measurement $\{Q_j, \overline Q_j\}$ requires an accurate estimate of the ground-state energy and spectral gap of $H_j$. While these could potentially be obtained from the quantum algorithm itself with some extra cost, it is reasonable to assume that, for intermediate size simulations, classical methods will be sufficiently powerful to provide an accurate solution to the static quantum problem. 

Details on the simulation method and the parameters used are provided in the appendix. 

\section{Numerical results} 

To obtain a general estimate of the resource scaling, we applied the numerical method outlined above to the Hubbard model on systems of various shapes and sizes, ranging from 2 to 65 sites. Some shapes, such as a square system, have additional ground-state degeneracies because of discrete symmetries. For simplicity, we explicitly discarded such systems. The problem could be circumvented by using a symmetry-breaking field~\cite{WHWC15a}, or by using an initial state in a fixed symmetry sector and using a Hamiltonian sequence that preserves this symmetry.  

In all cases, the initial Hamiltonian $H_0$ was obtained by turning off the Coulomb interaction $U$. The corresponding ground state is a fermionic Gaussian state which can be easily prepared on a quantum computer~\cite{PHWW14a}. 

To adiabatically reach the final Hamiltonian, we perform a linear interpolation. It has been shown that this is generally not an optimal path~\cite{WHWC15a}, but it has the advantage of simplicity and universality. 

\begin{figure}
\centering
\includegraphics[width=8.6cm]{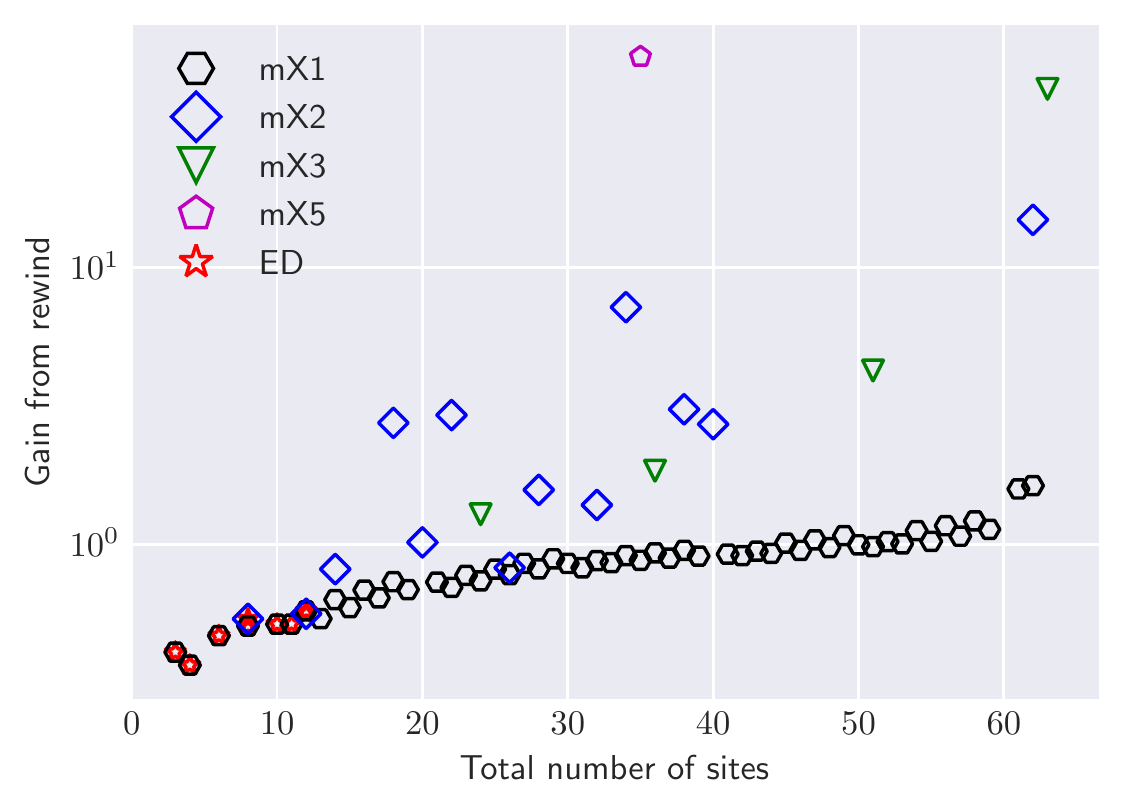}
\caption{Gain obtained from the rewind procedure to prepare the ground state of the 10\% doped Hubbard model on $m \times k$ lattices. The units of the TTS are set by the natural units of the Hubbard model ($T_\text{hub}=1$).}
\label{fig:Gain}
\end{figure}

\begin{figure}
\centering
\includegraphics[width=8.6cm]{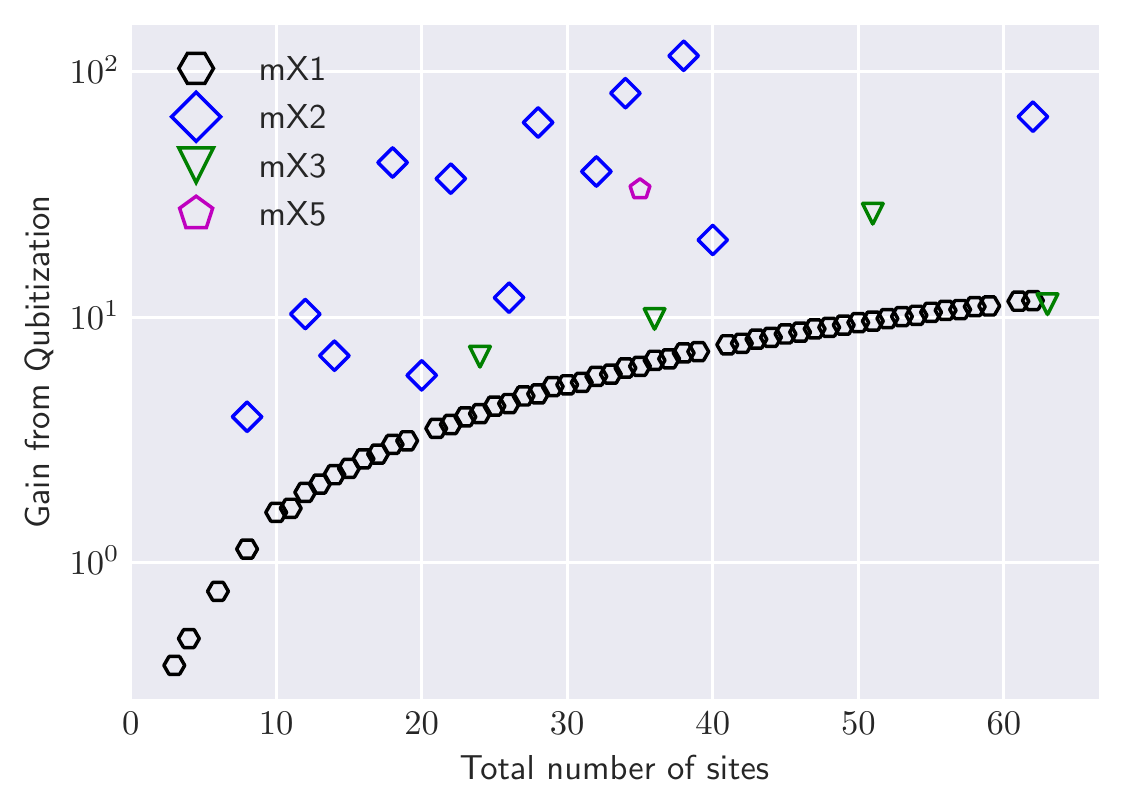}
\caption{Gain obtained from qubitization with rewind compared to a direct rewind procedure. }
\label{fig:gainQub}
\end{figure}

The gain is defined as the ratio of the minimum TTS of the first method to the minimum TTS of the second method. The improvement brought by the rewind procedure alone is shown in Fig.~\ref{fig:Gain} and the gain when added to qubitization is shown in Fig.~\ref{fig:gainQub}. We observe that for small systems, the results obtained with DMRG is in near perfect agreement with exact diagonalization. This leads us to trust that it provides reliable estimates for larger systems.

With some exceptions, the rewind procedure for small gap systems (smaller than $0.1$) offers a gain of about one order of magnitude. These systems include chains of about 50-60 sites and 2D systems of 20-60 sites. While our simulations are too limited to extract a clear trend, the improvement appears to increase with lattice size--or, equivalently, with the inverse gap. The rewind procedure will thus become crucial, already for intermediate-size quantum simulation algorithms. 

The qubitization introduced in~\cite{DSHT17a} offers a gain as large as $10^2$ over the rewind procedure, i.e., $10^3$ over the standard discrete adiabatic state preparation algorithm. Since this method can efficiently prepare the eigenstates of its walk operator, it can greatly reduce the circuit depth. 

The original adiabatic algorithm is known to have a scaling bounded by $1/\Delta ^ 2$~\cite{959902}. However, in the formulas introduced for our TTS computation, Eqs.(\ref{eq:TTS1}) and (\ref{eq:TTS2}), only a factor of $1/\Delta$ is explicit. Nevertheless, the fidelity term has an implicit dependence on $\Delta$ related to the density of steps in the discretization. Because we use a TTS approach to define the schedule, we expect the contribution of that term to be at most $1/{\Delta}$, but it is likely faster for each instance. The scaling improvement given by the faster schedule in our current application is studied in the appendix.

To obtain the circuit complexity, we need to quantify the number of elementary gates required to compute the operator $U_j = e^{-iH_j}$. There are different ways to approximate time-evolution operator $U(t)=e^{-iHt}$ as a quantum circuit, resulting in different gate counts, that have recently been estimated in Ref.~\cite{CMNR18a}. In the most efficient case, the simulation of a 100-qubit system for a time $t=100$ requires $10^9$ $T$ gates. Since these product-formula algorithms can be parallelized, this implies a $10^7$ circuit depth. So to produce a time-evolution operator $U(t)$ with $t\sim 10^5-10^7$ requires a quantum circuit of depth $10^{12}-10^{14}$ (ignoring logarithmic corrections). With a microsecond logical gate time, the quantum computation time would be from a week to a year long.  

If our ground-state measurement uses the unitary walk operator of Ref.~\cite{DSHT17a} instead of the time-evolution operator, then the spectral gap $\Delta$ in formulas (\ref{eq:TTS1})-(\ref{eq:TTS2}) should be replaced by $\arccos(1-\Delta/\mathcal{N})$. We find that this method requires $10^4-10^{6}$ applications of the unitary walk operator instead of the $10^5-10^7$ unit-time evolution for systems with $N \approx 100$ sites. The circuit depth for implementing a single walk operator to precision $\epsilon$ can be compressed~\cite{LHPS19a} to $3\log(N)\log\frac 1\epsilon$, where $\epsilon$ is the accuracy with which logical gates are synthesized. An accuracy $\epsilon = \sqrt{\Delta}/(100 N^2)$ should do~\cite{DSHT17a}, resulting in a circuit depth $10^5-10^8$ for $N \approx 100$.

\begin{figure}
\centering
\includegraphics[width=8.6cm]{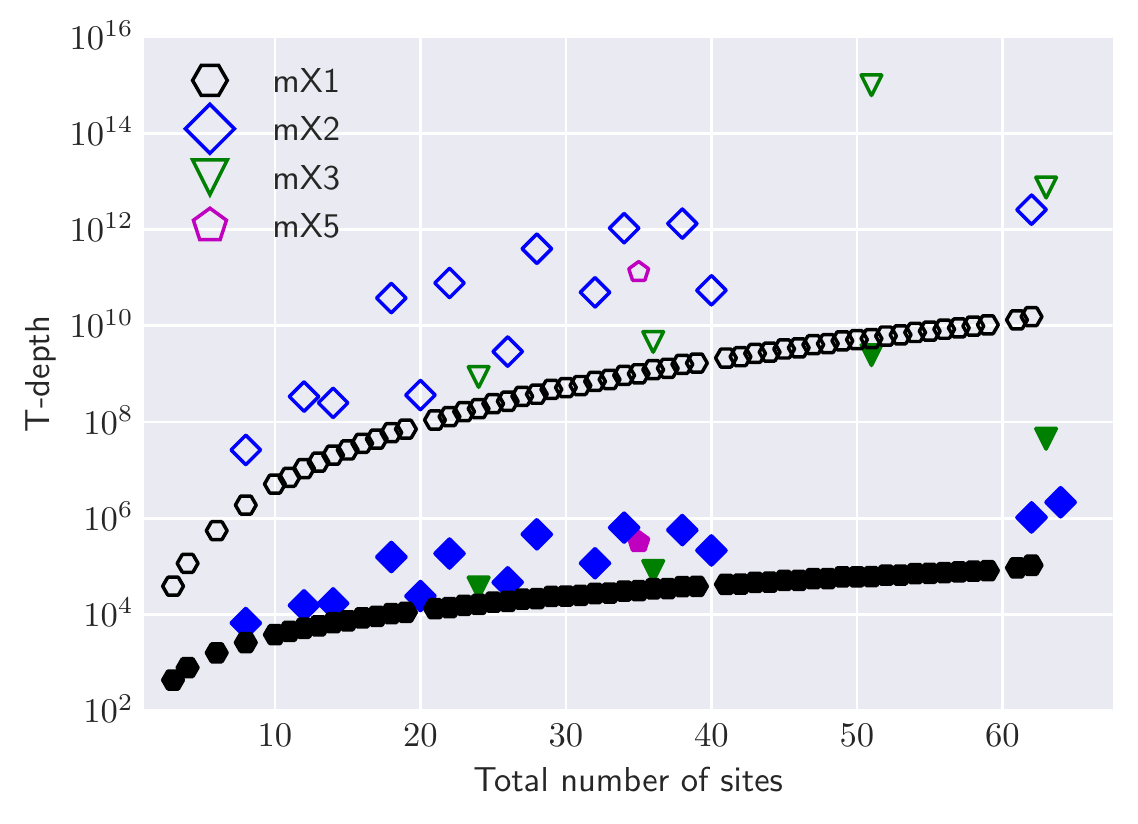}
\caption{ $T$-depth of the circuit presented is this work. Discrete adiabatic evolution with qubitization and rewind procedure (filled symbols) vs high-order products using sixth-order formulas with empirical error bound~\cite{CMNR18a} and rewind procedures only (empty symbols). }
\label{fig:t-depth}
\end{figure} 

These circuit complexities are shown in Fig.~\ref{fig:t-depth}, which compares the $T$-depth of the rewind procedure using product formulas simulated in Ref.~\cite{CMNR18a} with all procedures described in this work. For systems of intermediate sizes, the gain is between one and six orders of magnitude.




\section{Conclusion} 

Building on well-established quantum many-body methods for classical computers, we have devised a procedure to estimate the cost of quantum ground-state preparation on quantum computers. Besides its use as a benchmark, our procedure could assist in the design and optimization of the quantum simulation algorithm. 

Our results show a significant improvement in required resources over the theoretical adiabatic bound. Indeed, qubitization lowers the implementation cost of the unitary used and the precision needed in quantum phase estimation. The latter changes part of the scaling from $1/{\Delta}$ to $1/\acos(1-{\Delta}/\mathcal{N})$, which corresponds to a gain for the systems studied in this work. 

By extrapolating the general trend of our results, we can predict that a time evolution as long as $10^6-10^{14}$ (in T-depth) is required to prepare the ground state of an intermediate-size Hubbard system with the adiabatic algorithm. We have also proposed an improved adiabatic optimization that decreases these times to the range $10^4-10^8$. The overall gain can be as large as $10^6$ for intermediate-size lattices.

These drops in computing time could be even more impressive when combined to a clever adiabatic path. For example, one could create a nonlinear interpolation, by using a symmetry-breaking field~\cite{WHWC15a}, as mentioned before. We expect that combining the latter with the method described in this paper would make an efficient algorithm for intermediate size error-corrected quantum computers. 

\begin{acknowledgments}
We thank Matthias Troyer for their suggestions, and Anirban Narayan Chowdhury and Thomas Baker for stimulating discussions. JL acknowledges support from the NSERC Canada Graduate Scholarships and the FRQNT programs of scholarships. Computing resources were provided by Compute Canada and Calcul Qu\'ebec. 
\end{acknowledgments}

\appendix

\section{Simulation method}
In this section, we provide details on the simulation methods used to solve the eigenproblem of the Hubbard models of different sizes. For small systems, where this is possible, we use exact diagonalization; otherwise we use DMRG to find an approximate solution. The goal of this section is to enable reproduction of the results.
\subsection{Exact diagonalization}
The exact results are computed with the Lanczos algorithm~\cite{lanczos1950iteration}. To reduce the matrix size, which scales as $4^N$ where $N$ is the number of sites, we only consider the subspace of interest where the numbers of up ($N_\uparrow$) and down ($N_\uparrow$) spins are fixed, leading to a matrix size 
\begin{align}
	D= \bigg(\frac{N!}{N_\uparrow !(N-N_\uparrow)!}\bigg)\bigg(\frac{N!}{N_\downarrow !(N-N_\downarrow)!}\bigg).
\end{align}
We chose the number of fermions to represent the worst-case scenario, i.e., the smallest gap. For the small lattices, we found that the smallest gap of the  Hubbard model is at roughly 10\% doping above (or equivalently below) half filling. We choose to keep this filling factor in the hope to be close to the smallest gap (at fixed number of spins) which correspond to hard problems in the adiabatic setting. For example, for a 64-sites lattice, half filling means 32 up spins and 32 down spins. At 10\% doping above (resp. below) half filling, we will have 35 or 36 (resp. 29 or 28) up and down spins. 
\subsection{Density-Matrix Renormalization Group}
For bigger lattice sizes,  exact diagonalization is intractable. We used DMRG, a tensor network method, to approximately solve the eigenproblem. See Ref.~\cite{S11b} for a review. The method has been implemented using the ITensor C++ library~\cite{ITensor}. To preserve the number of up and down spins, we used the IQTensor object.

\begin{figure}
	\centering
	\includegraphics[width=8.6cm]{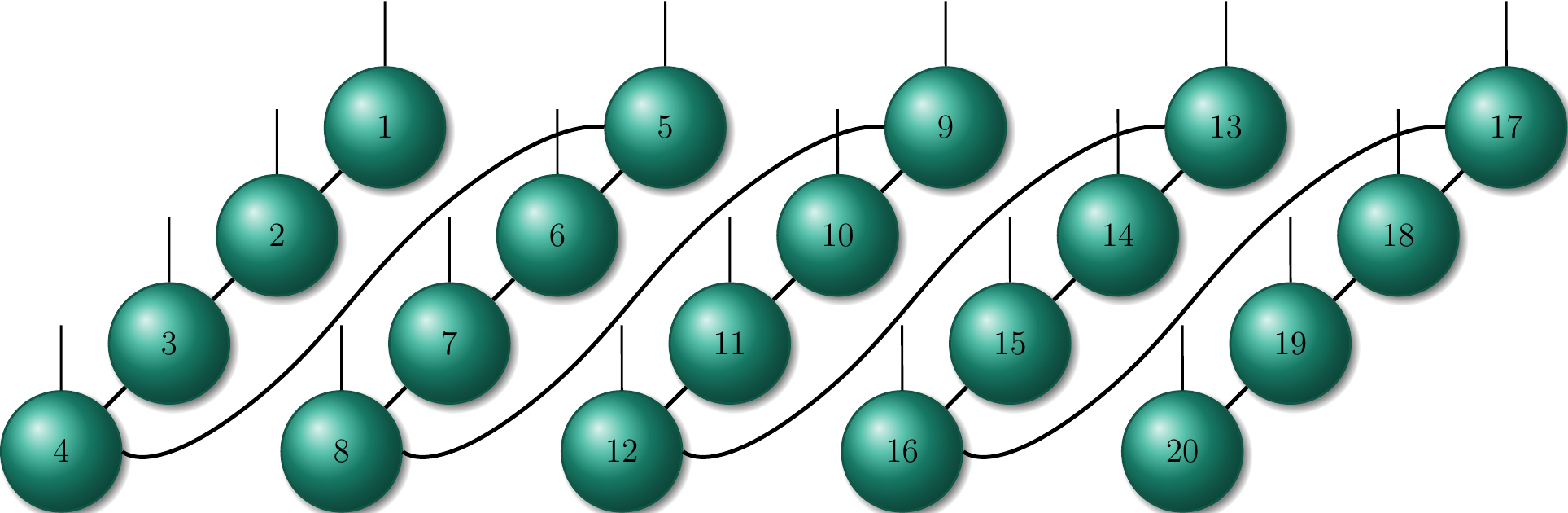}
	\caption{MPS of the $m \times k $ lattice ($m=5$,$k=4$)}
	\label{fig:mps4-5}
\end{figure}

The $m\times k$ lattices (for a total of $k m = N$ sites and $m ­> k$) are mapped to a MPS (see Fig.~\ref{fig:mps4-5}). 

\begin{table}
\centering
\begin{tabular}{ c | c | c | c | c } 
 maxm & minm & cutoff & niter & noise \\
 \hline 
 2 & 1 & $10^{-5}$ & 2 & $10^{-1}$ \\ 
 2 & 1 & $10^{-6}$ & 2 & $10^{-1}$ \\ 
 4 & 1 & $10^{-7}$ & 2 & $10^{-2}$ \\
 4 & 1 & $10^{-8}$ & 2 & $10^{-2}$ \\ 
 8 & 1 & $10^{-8}$ & 2 & $10^{-3}$ \\ 
 8 & 1 & $10^{-8}$ & 2 & $10^{-3}$ \\
 16 & 1 & $10^{-8}$ & 2 & $10^{-4}$ \\
 16 & 1 & $10^{-9}$ & 2 & $10^{-4}$ \\
 32 & 1 & $10^{-9}$ & 2 & $10^{-5}$ \\
 32 & 1 & $10^{-10}$ & 2 & $10^{-5}$ \\
 64 & 1 & $10^{-10}$ & 2 & $10^{-6}$ \\
 64 & 1 & $10^{-10}$ & 2 & $10^{-6}$ \\
 64 & 1 & $10^{-10}$ & 2 & $10^{-6}$ \\
 128 & 1 & $10^{-10}$ & 2 & $10^{-7}$ \\
 128 & 1 & $10^{-10}$ & 2 & $10^{-7}$ \\
 256 & 1 & $10^{-10}$ & 2 & $10^{-7}$ \\
 256 & 1 & $10^{-10}$ & 2 & $10^{-8}$ \\
 512 & 1 & $10^{-10}$ & 2 & $10^{-8}$ \\
 512 & 1 & $10^{-10}$ & 2 & $10^{-8}$ \\
 1024 & 1 & $10^{-10}$ & 2 & $10^{-9}$ \\
\end{tabular}
\caption{the schedule of the sweep parameter given to the IQTensor DMRG function (maximal bound dimension -- maxm, minimal bound dimension -- minm, the truncation error cutoff -- cutoff, the number of Davidson iterations -- niter and the noise term -- noise). The last row is the parameter used for the remaining sweeps. }
\label{tab:input}
\end{table}

The energy error goal (tolerated error on the energy) is set to $10^{-9}$, the number of sweep is set to 30 to reach the ground state and to 40 to reach the first-excited state. Note that to find the first-excited state, we need to set a penalty $\tilde{\Delta} \in \mathbb{R}_+^*$ to the ground state by running the DMRG again for the following Hamiltonian :
\begin{align}
H' = H + \tilde{\Delta} \ketbra{\psi_0}
\end{align}
where $\ket{\psi_0}$ is the ground state of $H$. By a trial and error approach, we found that setting $\tilde{\Delta}$ to the absolute value of the ground-state energy if that value is higher than 1, and to 10 if it is smaller than or equal to 1 give better result and convergence rate. 

For the other ITensor parameters, we built a ramp to avoid getting stuck in a local minimum (see Table~\ref{tab:input}).

\section{Scaling}

The scaling of the quantum adiabatic algorithm is bounded above by $O\bigg(\frac{1}{\Delta^2}\bigg)$~\cite{959902}, where $\Delta$ is the minimal gap of the time-dependent Hamiltonian 
\begin{align}
  H(t) = \bigg(1- \frac{t}{\tau} \bigg)H_0 + \frac{t}{\tau} H_f.
\end{align}
$\tau$ is the adiabatic parameter, $H_0$ is the initial Hamiltonian with an easy-to-prepare ground state and $H_f$ is the final Hamiltonian with the target ground state. In the discrete version, the equivalent of the delay schedule $\tau$ is defined by the density of steps and the precision of the energy measurements. 

One could use a faster schedule in some instances and obtain a better scaling than the theoretical bound. The procedure introduced in the main paper results in a faster schedule than constant-step linear interpolation. The minimal TTS finds a trade-off between a long high-probability computation and short low-probability computation repeated multiple times. We expect the scaling corresponding to the density of step to be at worst $1/\Delta$ for all instances. 

\begin{figure}
\centering
\includegraphics[width=8.6cm]{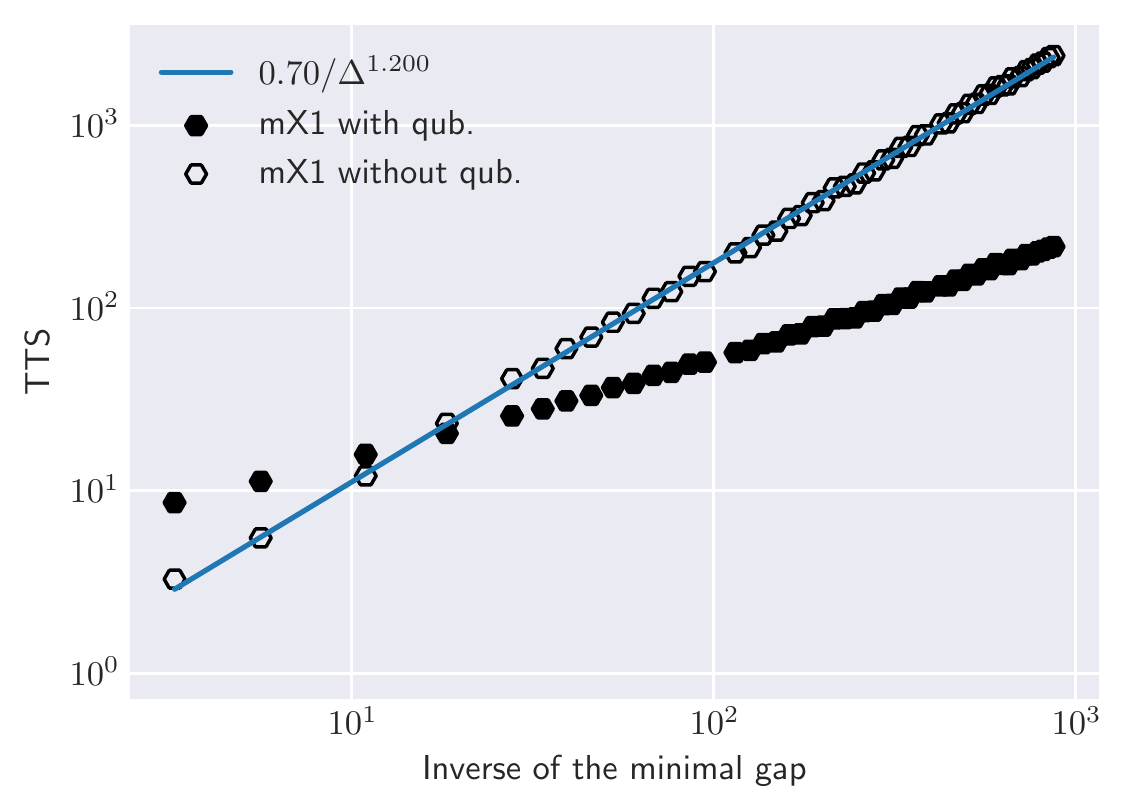}
\caption{ TTS obtained with and without qubitization (empty and filled symbols reps.) on chain, $m \times 1$, lattices. The units of the TTS are set by the natural units of the Hubbard model ($T_\text{hub}=1$). The $x$-axis corresponds to the minimal gap of the original problem also in the natural units of the Hubbard model (before qubitization).  }
\label{fig:mx1}
\end{figure}

By using qubitization, we modified both the gap and states used for the evolution. The change in the gap has an obvious impact on the scaling, changing a factor from $1/\Delta$ to $1/\acos(1-\bar \Delta)$, where $\bar\Delta$ is the gap of the normalized Hamiltonian. Qubitization will not change the density of steps used but will affect the success probability of each step since the states used are the eigenstates of $W_j=\exp{i \acos(\bar H_j )}$. 

\begin{figure}[h]
\centering
\includegraphics[width=8.6cm]{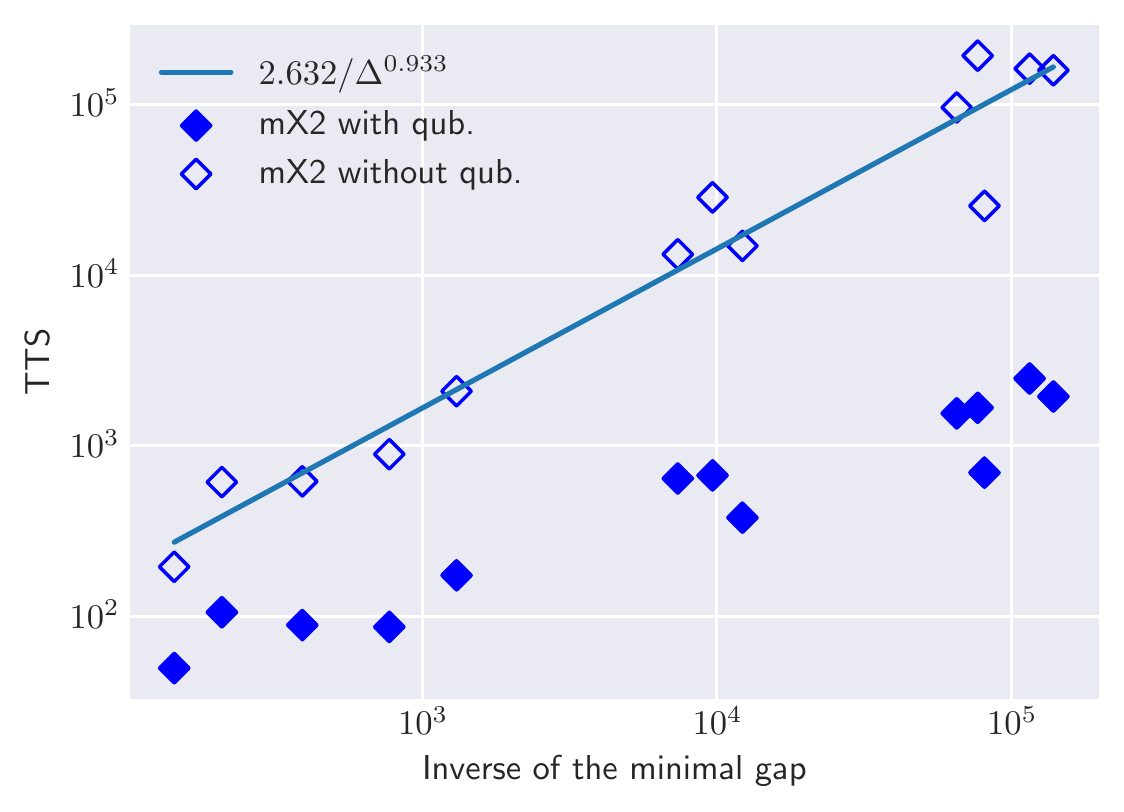}
\caption{ TTS obtained with and without qubitization (empty and filled symbols reps.) on ladder, $m \times 2$, lattices. The units of the TTS are set by the natural units of the Hubbard model ($T_\text{hub}=1$). The $x$-axis corresponds to the minimal gap of the original problem also in the natural units of the Hubbard model (before qubitization). }
\label{fig:mx2}
\end{figure}

An exact study of the scaling including all contributions (qubitization, schedule choice, rewind protocol) is hard to perform. To get a sense of the scaling of our simulation, see Figs.~\ref{fig:mx1} and~\ref{fig:mx2}. Because the simulation results appear to depend on lattice depth, we focused on ladder and chain lattices ($m \times 2$ and $m \times 1$). The horizontal axis is the inverse minimal gap of the initial problem, thus, before qubitization. Since the qubitization increases the gap size, it also reduces greatly the cost of the adiabatic evolution (filled symbols). The scaling is better than $1/\Delta^{2}$ in all cases we studied. The density of steps should be bounded by $1/\Delta$. However, the different trends from the different lattice shapes indicate that the gap is not the only determinant factor in our scaling. The implicit term in the TTS is nontrivial. From the numerics, there is a significant speedup, but we cannot distinguish what is general from what is specific to the system used. Thus, the fitting curves on the figures should be considered as guides, not as an actual scaling of the method. 

\end{document}